\documentclass[prl,aps,twocolumn,superscriptaddress,showpacs,preprintnumbers,
               amsmath,amssymb,floatfix]{revtex4}

\def\gtwid{\mathrel{\raise.3ex\hbox{$>$\kern-.75em\lower1ex\hbox{$\sim$}}}}
\def\ltwid{\mathrel{\raise.3ex\hbox{$<$\kern-.75em\lower1ex\hbox{$\sim$}}}}

\usepackage{graphicx}
\usepackage{dcolumn}
\usepackage{bm}

\def\gev{GeV/c$^2$}

\begin{document}


\title{Limits on spin-independent WIMP-nucleon interactions from the 
two-tower run of the Cryogenic Dark Matter Search}


\affiliation{Department of Physics, Brown University,
Providence, RI 02912, USA}
\affiliation{Department of Physics, Case Western
Reserve University, Cleveland, OH  44106, USA}
\affiliation{Fermi National Accelerator Laboratory,
Batavia, IL 60510, USA}
\affiliation{Lawrence Berkeley National Laboratory,
Berkeley, CA 94720, USA}
\affiliation{Department of Physics, Santa Clara University, Santa
Clara, CA 95053, USA}
\affiliation{School of Physics \& Astronomy, University of Minnesota,
Minneapolis, MN 55455, USA}
\affiliation{Department of Physics, Stanford University,
Stanford, CA 94305, USA}
\affiliation{Department of Physics, University of
California, Berkeley, CA 94720, USA}
\affiliation{Department of Physics, University of
California, Santa Barbara, CA 93106, USA}
\affiliation{Department of Physics, University of
Colorado at Denver and Health Sciences Center, Denver, CO 80217, USA}

\author{D.S.~Akerib} \affiliation{Department of Physics, Case Western Reserve University, Cleveland, OH  44106, USA} 
\author{M.J.~Attisha} \affiliation{Department of Physics, Brown University, Providence, RI 02912, USA} 
\author{C.N.~Bailey} \affiliation{Department of Physics, Case Western Reserve University, Cleveland, OH  44106, USA} 
\author{L.~Baudis} \affiliation{Department of Physics, University of Florida, Gainesville, FL 32611, USA} 
\author{D.A.~Bauer} \affiliation{Fermi National Accelerator Laboratory, Batavia, IL 60510, USA} 
\author{P.L.~Brink} \affiliation{Department of Physics, Stanford University, Stanford, CA 94305, USA} 
\author{P.P.~Brusov} \affiliation{Department of Physics, Case Western Reserve University, Cleveland, OH  44106, USA} 
\author{R.~Bunker} \affiliation{Department of Physics, University of California, Santa Barbara, CA 93106, USA} 
\author{B.~Cabrera} \affiliation{Department of Physics, Stanford University, Stanford, CA 94305, USA} 
\author{D.O.~Caldwell} \affiliation{Department of Physics, University of California, Santa Barbara, CA 93106, USA} 
\author{C.L.~Chang} \affiliation{Department of Physics, Stanford University, Stanford, CA 94305, USA} 
\author{J.~Cooley} \affiliation{Department of Physics, Stanford University, Stanford, CA 94305, USA} 
\author{M.B.~Crisler} \affiliation{Fermi National Accelerator Laboratory, Batavia, IL 60510, USA} 
\author{P.~Cushman} \affiliation{School of Physics \& Astronomy, University of Minnesota, Minneapolis, MN 55455, USA} 
\author{M.~Daal} \affiliation{Department of Physics, University of California, Berkeley, CA 94720, USA} 
\author{R.~Dixon} \affiliation{Fermi National Accelerator Laboratory, Batavia, IL 60510, USA} 
\author{M.R.~Dragowsky} \affiliation{Department of Physics, Case Western Reserve University, Cleveland, OH  44106, USA} 
\author{D.D.~Driscoll} \affiliation{Department of Physics, Case Western Reserve University, Cleveland, OH  44106, USA} 
\author{L.~Duong} \affiliation{School of Physics \& Astronomy, University of Minnesota, Minneapolis, MN 55455, USA} 
\author{R.~Ferril} \affiliation{Department of Physics, University of California, Santa Barbara, CA 93106, USA} 
\author{J.~Filippini} \affiliation{Department of Physics, University of California, Berkeley, CA 94720, USA} 
\author{R.J.~Gaitskell} \affiliation{Department of Physics, Brown University, Providence, RI 02912, USA} 
\author{S.R.~Golwala} \affiliation{Department of Physics, California Institute of Technology, Pasadena, CA 91125, USA} 
\author{D.R.~Grant} \affiliation{Department of Physics, Case Western Reserve University, Cleveland, OH  44106, USA} 
\author{R.~Hennings-Yeomans} \affiliation{Department of Physics, Case Western Reserve University, Cleveland, OH  44106, USA} 
\author{D.~Holmgren} \affiliation{Fermi National Accelerator Laboratory, Batavia, IL 60510, USA} 
\author{M.E.~Huber} \affiliation{Department of Physics, University of Colorado at Denver and Health Sciences Center, Denver, CO 80217, USA} 
\author{S.~Kamat} \affiliation{Department of Physics, Case Western Reserve University, Cleveland, OH  44106, USA} 
\author{S.~Leclercq} \affiliation{Department of Physics, University of Florida, Gainesville, FL 32611, USA} 
\author{A.~Lu} \affiliation{Department of Physics, University of California, Berkeley, CA 94720, USA} 
\author{R.~Mahapatra} \affiliation{Department of Physics, University of California, Santa Barbara, CA 93106, USA} 
\author{V.~Mandic} \affiliation{Department of Physics, University of California, Berkeley, CA 94720, USA} 
\author{P.~Meunier} \affiliation{Department of Physics, University of California, Berkeley, CA 94720, USA} 
\author{N.~Mirabolfathi} \affiliation{Department of Physics, University of California, Berkeley, CA 94720, USA} 
\author{H.~Nelson} \affiliation{Department of Physics, University of California, Santa Barbara, CA 93106, USA} 
\author{R.~Nelson} \affiliation{Department of Physics, University of California, Santa Barbara, CA 93106, USA} 
\author{R.W.~Ogburn} \affiliation{Department of Physics, Stanford University, Stanford, CA 94305, USA} 
\author{T.A.~Perera} \affiliation{Department of Physics, Case Western Reserve University, Cleveland, OH  44106, USA} 
\author{M.~Pyle} \affiliation{Department of Physics, Stanford University, Stanford, CA 94305, USA}
\author{E.~Ramberg} \affiliation{Fermi National Accelerator Laboratory, Batavia, IL 60510, USA} 
\author{W.~Rau} \affiliation{Department of Physics, University of California, Berkeley, CA 94720, USA}
\author{A.~Reisetter} \affiliation{School of Physics \& Astronomy, University of Minnesota, Minneapolis, MN 55455, USA} 
\author{R.R.~Ross} \thanks{Deceased} \affiliation{Lawrence Berkeley National Laboratory, Berkeley, CA 94720, USA} \affiliation{Department of Physics, University of California, Berkeley, CA 94720, USA}
\author{B.~Sadoulet} \affiliation{Lawrence Berkeley National Laboratory, Berkeley, CA 94720, USA} \affiliation{Department of Physics, University of California, Berkeley, CA 94720, USA}
\author{J.~Sander} \affiliation{Department of Physics, University of California, Santa Barbara, CA 93106, USA} 
\author{C.~Savage} \affiliation{Department of Physics, University of California, Santa Barbara, CA 93106, USA} 
\author{R.W.~Schnee} \affiliation{Department of Physics, Case Western Reserve University, Cleveland, OH  44106, USA} 
\author{D.N.~Seitz} \affiliation{Department of Physics, University of California, Berkeley, CA 94720, USA} 
\author{B.~Serfass} \affiliation{Department of Physics, University of California, Berkeley, CA 94720, USA} 
\author{K.M.~Sundqvist} \affiliation{Department of Physics, University of California, Berkeley, CA 94720, USA} 
\author{J-P.F.~Thompson} \affiliation{Department of Physics, Brown University, Providence, RI 02912, USA} 
\author{G.~Wang} \affiliation{Department of Physics, California Institute of Technology, Pasadena, CA 91125, USA} \affiliation{Department of Physics, Case Western Reserve University, Cleveland, OH  44106, USA}
\author{S.~Yellin} \affiliation{Department of Physics, Stanford University, Stanford, CA 94305, USA} \affiliation{Department of Physics, University of California, Santa Barbara, CA 93106, USA}
\author{J.~Yoo} \affiliation{Fermi National Accelerator Laboratory, Batavia, IL 60510, USA} 
\author{B.A.~Young} \affiliation{Department of Physics, Santa Clara University, Santa Clara, CA 95053, USA}

\collaboration{CDMS Collaboration}

\noaffiliation


\begin{abstract}
We report new results from the Cryogenic Dark Matter Search (CDMS II)
at the Soudan Underground Laboratory.  Two towers, each consisting of
six detectors, were operated for 74.5 live days, giving
spectrum-weighted exposures of
34\,kg-d for germanium and 12\,kg-d for silicon targets after cuts,
averaged over recoil energies 10--100\,keV for a WIMP mass of 60\,\gev.
A blind analysis was
conducted, incorporating improved techniques for rejecting surface events. No WIMP signal exceeding expected
backgrounds was observed.  When combined with our previous results
from Soudan, the 90\% C.L. upper limit on the spin-independent
WIMP-nucleon cross section is 1.6$\times10^{-43}$\,cm$^2$ from Ge, and
3$\times10^{-42}$\,cm$^2$ from Si, for a WIMP mass of 60\,\gev.
The combined limit from Ge (Si) is a factor of 2.5 (10) lower than our previous results, and constrains predictions of supersymmetric models.
\end{abstract}


\maketitle

One quarter of the energy density of the universe
consists of non-baryonic dark matter \cite{Cosmol}, which is
gravitationally clustered
in halos surrounding visible galaxies~\cite{Galax}. 
The Weakly Interacting
Massive Particle (WIMP) \cite{WIMP}, 
a dark matter candidate, arises independently from
considerations of Big Bang cosmology and from
supersymmetric phenomenology, where the neutralino
can be a WIMP~\cite{BigBPart,Reviews}.  
A WIMP has a scattering cross section with an atomic
nucleus characteristic of the weak interaction and a mass
comparable to that of an atomic nucleus.

The CDMS II experiment \cite{akerib04, prd118} is designed to detect
recoils of 
atomic nuclei that have been scattered by incident WIMPs in germanium
(Ge) and silicon (Si) crystals.  Events with recoil energies of a few
tens of keV and rates $\lesssim$1\,event\,kg$^{-1}$\,d$^{-1}$ are
expected \cite{Reviews, lewin}.  The CDMS II search is most sensitive
to spin-independent (SI) WIMP-nucleon scattering amplitudes
\cite{SpinInt}.  Coherent superposition of SI amplitudes gives Ge better sensitivity than Si,
except for small WIMP masses, where kinematics increase the Si
sensitivity.  In this report we interpret our data with the SI ansatz.
Another report describes a spin-dependent interpretation of our data
\cite{SDPRL}.

The CDMS II apparatus in the Soudan Underground Laboratory has been
described elsewhere \cite{akerib04, prd118}.  At the experiment's core,
Z(depth)-sensitive Ionization and Phonon detectors (ZIPs)
measure the ionization and athermal phonon signals caused by
recoiling particles in Ge and Si crystals~\cite{zips, prd118}.  We
report new results from the most recent CDMS data run 
collected between March 25 and August 8, 2004,
consisting of 74.5 live days.
Six Ge (250\,g each) and six Si (100\,g each)
ZIPs were operated in two vertical stacks (``towers'') at a temperature
of 50\,mK.  This report includes the first data from Tower 2, 
which contains four Si and two Ge ZIPs. Improvements made since the 
previous run~\cite{akerib04, prd118} include a reduction of electrical 
noise and more frequent infrared illumination
to clear the crystals of space-trapped charge.

The ZIP detector provides event-by-event discrimination of nuclear
recoils from the dominant background of electron recoils.  The ratio
of ionization energy to phonon energy (``ionization yield'') is
$\sim$~0.3 for nuclear recoils, normalized to electron recoils (see
yield$\sim$1 in Fig.~1).  Electron recoils near the detector surface
suffer from poor ionization collection and can mimic the ionization yield
of nuclear recoils that occur throughout the detector.  
Each ZIP's phonon sensors are divided into four quadrants.  
Timing and signal
amplitude comparisons among the quadrants provide discrimination
against electron recoils, particularly those near the surface.
Also, surface electron recoils often deposit energy in adjacent ZIPs
within a tower, causing multiple-detector events. Energy from a WIMP
recoil would be contained in one detector.

To avoid bias, we performed blind analyses.
Events in and near the signal region were removed from WIMP-search
data sets, or ``masked'' \cite{softbug}.
The cuts that define a signal were
determined using calibration data from $^{133}$Ba and $^{252}$Cf
sources and from the non-masked WIMP-search data. 
Seven million electron recoils were collected using a $^{133}$Ba
source of gamma rays, exceeding the comparable WIMP-search data
by a factor of twenty.  Half of the $^{133}$Ba data were used
to define analysis cuts and the other half to test
cut definitions and estimate expected backgrounds. 
The detector response to nuclear recoils
was characterized with 10$^4$ neutrons from a $^{252}$Cf
source, collected in four separate, several-hour periods during the run.  

Data from the $^{133}$Ba source were used to monitor detector stability
and to characterize detector performance.  One Ge detector
in Tower 2, ZIP 5 (T2Z5(Ge)), had a spatial region of abnormal ionization
response which we excluded from analysis.
The Si detector T1Z6, known to be contaminated with $^{14}$C, a
beta emitter, was excluded, as were detectors T1Z1(Ge) and
T2Z1(Si) due to poor phonon sensor performance.  We
report results from the remaining 5 Ge and 4 Si ZIPs, chosen before unmasking the WIMP signal region.

\begin{figure}[ht]
\begin{center}
\includegraphics[width=3.25in]{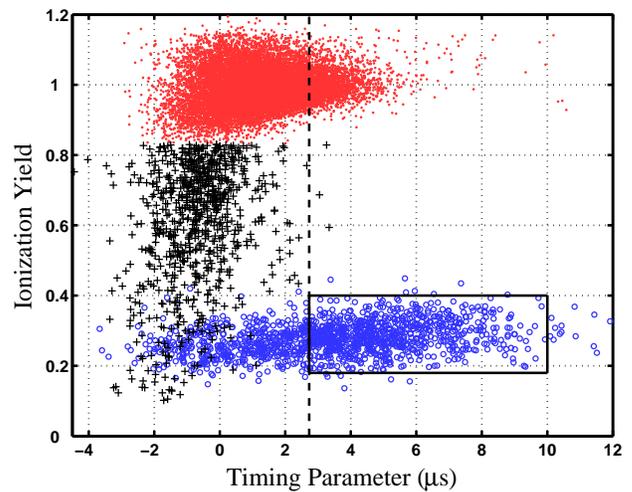} 
\caption{\small Ionization yield versus timing parameter (see text) for
calibration data in T2Z3(Ge), with recoil energies in the range
10--100\,keV.  Typical bulk-electron recoils 
from the $^{133}$Ba source of gamma rays are dots (red), with yield
near unity. Low-yield $^{133}$Ba events\,(+, black), 
attributed to surface electron
recoils, have small values of timing parameter, allowing discrimination
from neutron-induced nuclear recoils from $^{252}$Cf\,($\circ$, blue), which
show a wide range of timing parameter values.
The vertical dashed line shows the minimum timing parameter allowed
for WIMP candidates, and the box shows the approximate signal region,
which is in fact weakly energy dependent. (Color online.)}
\label{fig:timing}
\end{center}
\end{figure}

Most event selection criteria, based on quantities from the four phonon
and two ionization pulses from each detector, are similar to those
described in our previous reports \cite{akerib04, prd118}. However, we
completed five distinct analyses \cite{cdms_theses} focused on improving
existing methods and developing new techniques to reject surface electron
recoils.  Events with low ionization yield in the $^{133}$Ba calibration
data, which are from surface electron recoils, were used to develop
rejection criteria. All cuts were frozen prior to unmasking the signal region.

Phonon pulses from surface recoils are more prompt 
than those from recoils in the detector bulk.
Two timing quantities in the quadrant with the largest
phonon signal or ``local quadrant'' 
are particularly powerful: the time delay of the
phonon signal relative to the fast ionization signal, and the
phonon pulse risetime \cite{akerib04, clarke00}.
For the first and simplest of the five analysis techniques, the sum of delay and risetime forms a timing parameter, after energy corrections to delay and risetime are applied to achieve an energy-independent distribution.		
Figure~\ref{fig:timing} shows the ionization yield
versus this timing parameter for a typical Ge detector.
The $^{133}$Ba calibration events with yields of 0.1 to 0.8, identified as surface electron recoils, show a smaller timing parameter than most nuclear recoils induced by the $^{252}$Cf source.

\begin{figure}[ht]
\begin{center}
\includegraphics[width=3.25in]{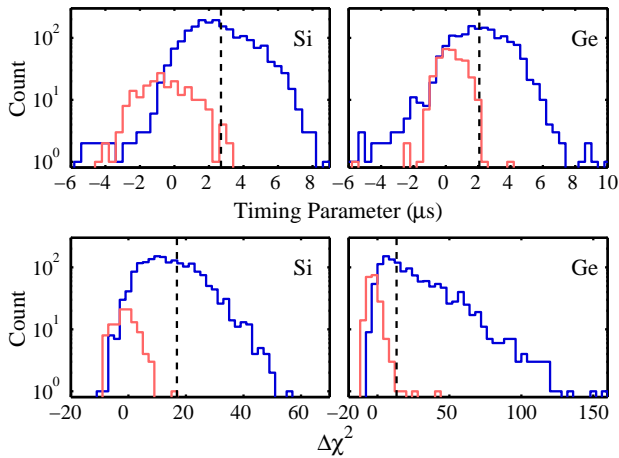}
\caption{\small Variables used to reject
surface electron recoils (7-100\,keV), for
data from T2Z4 (Si/left) and T1Z2 (Ge/right), for the timing
parameter (upper) and $\Delta\chi^2$ (lower).
Light (red) lines show distributions of
low-yield electron recoils from the $^{133}$Ba source, while
dark (blue) lines show distributions of nuclear recoils 
from the $^{252}$Cf source. Dashed lines indicate the minimum
values for acceptable WIMP candidates.  A cut on
the timing parameter is used for the Ge detectors, while
a requirement on $\Delta\chi^2$ is used for the Si detectors. (Color online.)}
\label{fig:sitiming}
\end{center}
\end{figure}

For data from the Ge detectors, we require that candidates for
WIMP-induced nuclear recoil exceed a minimum value for this timing
parameter (see Fig.~\ref{fig:sitiming} upper-right).  Because this
``timing parameter'' analysis is simple and robust, we agreed before
unmasking to report its results.  The expected sensitivities computed
for the four more advanced analyses described below were all within
$\pm$20\% of that computed for this technique.  

Two of the four advanced analysis techniques evaluate the goodness of fit,
$\chi^2$, for surface electron versus bulk nuclear recoil hypotheses.
These methods use three variables including their correlations: time
delay, risetime, and ``partition". The partition parameter, a measure
of energy distribution between the phonon quadrants, is defined as the
ratio of phonon energy in the local quadrant to that in the diagonal
quadrant.  The difference in $\chi^2$ between surface electron and
bulk nuclear recoil hypotheses, $\Delta\chi^2$, is used as the
principal discrimination parameter in these analyses. One method uses
energy-dependent variances and the other energy-independent variances.

For the Si detectors, we decided after unmasking to use the energy-dependent $\chi^2$ technique,
which has the best expected sensitivity to a nuclear recoil signal
from low-mass WIMPs for any of our five methods.  This analysis
technique is four times as sensitive as the timing parameter analysis 
for low-mass WIMPs.  Before
unmasking, we set the minimum requirement on the $\Delta\chi^2$ for
surface electron recoil rejection (see Fig.~\ref{fig:sitiming}
lower-left).  These criteria are used for our WIMP-search results for
Si.

The two remaining analysis techniques confirm the robustness of our Ge and Si results.  One technique combines delay,
risetime and partition in a neural net analysis, and the other
technique exploits additional information from the fitted signal
pulses to reconstruct recoil position and type.  These two
and the $\chi^2$ techniques promise further
improvements in sensitivity for the larger exposures planned in our
future runs, beyond the improvements already demonstrated in
Fig.~\ref{fig:sitiming}.


\begin{figure}[ht]
\begin{center}
\includegraphics[width=3.25in]{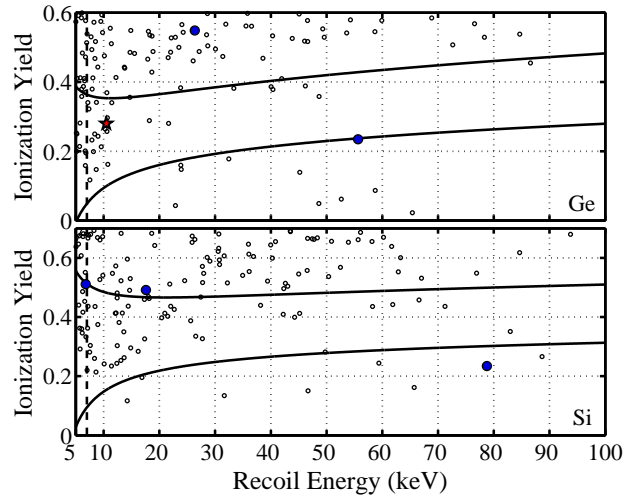}    
\caption{\small Ionization yield versus recoil energy for events in
all Ge detectors (upper) 
and all Si detectors (lower) passing initial data selection cuts prior to applying the surface electron recoil rejection cut.  
The signal region consists of recoil energies exceeding 7\,keV, shown with a
vertical dashed line, and yields between the curved lines defined with recoils 
induced by the $^{252}$Cf source while WIMP-search data were still masked.  
Below 7\,keV, separation between 
nuclear and electron recoils becomes poor.
Events passing the surface electron recoil cut are a star (red) 
inside the signal region 
and dark filled circles (blue) outside.
Bulk-electron recoils with yield near unity are above the vertical scale limits. (Color online.)}
\label{fig:leakage}
\end{center}
\end{figure}

Upon unmasking the Ge data, one candidate with 10.5\,keV recoil energy
was found to pass all cuts in the timing parameter analysis.  Unmasking the Si data revealed that no
events passed all cuts. Figure~\ref{fig:leakage} shows the unmasked data
in ionization yield versus recoil energy; the data before and after
application of the cut intended to reject surface electron recoils are
shown.  All analysis techniques showed
consistent results.

After unmasking the Ge data, we realized that the candidate
occurred in a detector during an interval of time when that
detector suffered inefficient ionization collection.
This defect by itself would prevent us from
claiming the event was evidence for a WIMP-induced nuclear recoil.
The candidate is also consistent with the rate of expected background.  
Although we do not claim this event as a nuclear recoil from a WIMP, we do include it in setting limits on the WIMP-nucleon cross section.

The expected surface-event backgrounds are estimated by multiplying
two factors.  The first factor is the number of events in the signal
region before surface-event cuts as obtained upon unmasking.  The second
factor is the fraction of surface events expected to pass the
surface-event cut.  This fraction may be estimated with the actual
passing fraction of low-yield events similar to the single-detector
event background.  An estimate using the passing fraction from
$^{133}$Ba Ge~(Si) data indicates a surface-event background of
0.13$\pm$0.05~(0.90$\pm$0.4) events.  However, we decided before
unmasking to base our background estimate on the passing fraction of
WIMP-search multiple-detector events in a wide low-yield region.  The
number of surface events expected to pass the surface-event cuts is
0.4$\pm$0.2\,(stat) $\pm$0.2\,(sys) between 10--100\,keV in Ge and
1.2$\pm$0.6\,(stat) $\pm$0.2\,(sys) between 7--100\,keV in Si.
Evidence suggests that beta decays of radioactive nuclides distributed
across the detectors cause most surface electron recoils in the
WIMP-search data.  The spatial distribution of these contaminants
differs from that of surface electron recoils from the $^{133}$Ba
source, which might explain the difference in estimated background
levels.  From simulation methods reported in \cite{prd118}, the
expected background due to cosmogenic neutrons that escape our muon
veto is 0.06 events in Ge and 0.05 events in Si.

\begin{figure}[ht]
\begin{center}
\includegraphics[width=3.25in]{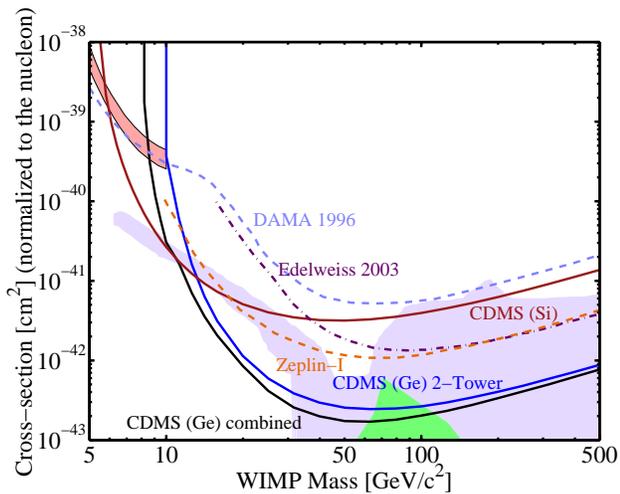}    
\caption{\small WIMP-nucleon cross section upper limits (90\% C.L.)
versus WIMP mass.  The upper CDMS Ge curve also uses data from the current
run, while the lower Ge curve includes the previous run
\cite{akerib04}.  Supersymmetric models allow the largest shaded (light-blue) region
\cite{bottino}, and the smaller shaded (green) region~\cite{ellis05}.  The shaded region in the upper left (see text) is from DAMA~\cite{gondolo}, and experimental limits are from DAMA~\cite{bernabei96}, EDELWEISS~\cite{sanglard05}, and ZEPLIN~\cite{zeplin}. (Color online.)}
\label{fig:limit}
\end{center}
\end{figure}

Figure~\ref{fig:limit} shows the upper limits on WIMP-nucleon cross
sections calculated from the Ge and Si analyses reported here using
standard assumptions for the galactic halo~\cite{lewin}.  For the
upper Ge limit, data between 10--100\,keV from this run are used.
Also shown is the combined limit obtained from this report and our
earlier work \cite{akerib04, prd118}.  For the combined Ge limit, we
have included data in the 7--10\,keV interval of recoil energy from
the run reported here \cite{extradata}.  The combined result for Ge
limits the WIMP-nucleon cross-section to $<1.6 \times
10^{-43}$\,cm$^2$ at the 90\% C.L. at a WIMP mass of 60\,\gev, a
factor of 2.5 below our previously published limits.  This new Ge
limit constrains some minimal supersymmetric (MSSM) parameter
space~\cite{bottino} and for the first time excludes some parameter
space relevant to constrained models (CMSSM)~\cite{ellis05}.

The Si limit in Fig.~\ref{fig:limit} is based on standard halo
assumptions using Si data from 7--100\,keV in this run.  
The Si result limits the
WIMP-nucleon cross-section to $<$3$\times10^{-42}$\,cm$^2$ at the
90\% C.L. at a WIMP mass of 60\,\gev.  This Si
result excludes new parameter space for low-mass WIMPs, including a
region compatible with interpretation of the DAMA signal (2--6 and
6--14\,keVee bins) as scattering
on Na~\cite{gondolo}.

This work is supported by the National Science Foundation under Grant
Nos. AST-9978911 and PHY-9722414, by the Department of Energy under contracts
DE-AC03-76SF00098, DE-FG03-90ER40569, DE-FG03-91ER40618, DE-FG02-94ER40823, and by
Fermilab, operated by the Universities Research Association, Inc.,
under Contract No. DE-AC02-76CH03000 with the Department of Energy.
The ZIP detectors were fabricated in the Stanford Nanofabrication
Facility operated under NSF.

%
%
%
%
%
%

\def\journal#1, #2, #3, #4#5#6#7{ 
  #1~{\bf #2}, #3 (#4#5#6#7)} 
\def\apl{\journal Appl.\ Phys.\ Lett., }
\def\apj{\journal Astrophys.\ J., }
\def\apjs{\journal Astrophys.\ J.\ Suppl., }
\def\app{\journal Astropart.\ Phys., }
\def\baas{\journal Bull.\ Am.\ Astron.\ Soc., }
\def\ejpc{\journal Eur.\ J.\ Phys.\ C., }
\def\lnp{\journal Lect.\ Notes\ Phys., }
\def\nature{\journal Nature, }
\def\nc{\journal Nuovo Cimento, }
\def\nima{\journal Nucl.\ Instr.\ Meth.\ A, }
\def\np{\journal Nucl.\ Phys., }
\def\npps{\journal Nucl.\ Phys.\ (Proc.\ Suppl.), }
\def\pl{\journal Phys.\ Lett., }
\def\prep{\journal Phys.\ Rep., }
\def\pr{\journal Phys.\ Rev., }
\def\prc{\journal Phys.\ Rev.\ C, }
\def\prd{\journal Phys.\ Rev.\ D, }
\def\prl{\journal Phys.\ Rev.\ Lett., }
\def\rsi{\journal Rev. Sci. Instr., }
\def\rpp{\journal Rep.\ Prog.\ Phys., }
\def\sjnp{\journal Sov.\ J.\ Nucl.\ Phys., }
\def\solarphys{\journal Solar Phys., }
\def\jetp{\journal J.\ Exp.\ Theor.\ Phys., }
\def\arnps{\journal Annu.\ Rev.\ Nucl.\ Part.\ Sci., }


\begin{thebibliography}{99}
\bibitem{Cosmol} D.~N.~Spergel {\it et al.},  (WMAP Collab.),
  \apjs 148, 175, 2003;
  M.~Tegmark {\it et al.},  (SDSS Collab.),
  \prd 69, 103501, 2004.
\bibitem{Galax}P.~Salucci and A.~Borriello, \lnp 616, 66, 2003;
  T.~Broadhurst {\it et al.}, \apj 621, 53, 2005.
\bibitem{WIMP} G.~Steigman and M.S.~Turner,
               \np B253, 375, 1985.
\bibitem{BigBPart} B.W.~Lee and S.~Weinberg, \prl 39, 165, 1977;
  S.~Weinberg, \prl 48, 1303, 1982.
\bibitem{Reviews}
  G.~Jungman, M.~Kamionkowski, and K.~Griest, \prep 267, 195, 1996;
  G.~Bertone, D.~Hooper, and J.~Silk, \prep 405, 279, 2005.   
\bibitem{akerib04} D.S.~Akerib {\it et al.}, (CDMS Collab.) \prl 93, 211301, 2004. 
\bibitem{prd118} D.S.~Akerib {\it et al.}, (CDMS Collab.), accepted for PRD.  arXiv:astro-ph/0507190 (2005).
\bibitem{lewin} J.D.~Lewin and P.F.~Smith, \app 6, 87, 1996.
\bibitem{SpinInt} A Majorana neutralino
undergoes a scalar interaction of roughly equal strength for neutrons
and protons, see K.~Griest \prd 38, 2357, 1988.
The SI scattering amplitude
is generally sensitive to scalar, vector, and tensor interactions of a spin-1/2
WIMP, see A.~Kurylov and M.~Kamionkowski, \prd 68, 103509, 2003.
\bibitem{SDPRL} D.S.~Akerib {\it et al.},  (CDMS Collab.), submitted to Phys.\ Rev.\ Lett.
\bibitem{zips}
K.D.~Irwin {\it et al.}, \rsi 66, 5322, 1995;
T.~Saab {\it et al.}, {\it AIP Proc.} {\bf 605}, 497 (2002).
\bibitem{softbug}Although software safeguards to enforce the
blinding scheme did not work as intended, the blinding policy itself
remained in effect, and we exercised care not to obtain any
information from inside the WIMP-search signal region.  We believe the
cuts were developed fully blind. 

\bibitem{cdms_theses} CDMS collaboration, in preparation for Phys. Rev. D; theses and notes at http://cdms.berkeley.edu/Dissertations.

\bibitem{clarke00} R.M.~Clarke {\it et al.}, \apl 76, 2958, 2000.


\bibitem{bottino} A.~Bottino 
 {\it et al.}, \prd 69, 037302, 2004.
\bibitem{bernabei96} R.~Bernabei {\it et al.}, Phys. Lett. B389, 757 (1996).
\bibitem{ellis05} J.~Ellis, 
{\it et al.}, \prd 71, 095007, 2005.
\bibitem{gondolo} We show the 90\% allowed region from Fig.~2a of 
P.~Gondolo and G.~Gelmini, \prd 71, 123520, 2005.
\bibitem{sanglard05} V.~Sanglard {\it et al.}, (EDELWEISS Collab.), \prd 71, 122002, 2005.
\bibitem{zeplin} G.J.~Alner {\it et al.}, (UK Dark Matter Collab.), \app 23, 444, 2005.
\bibitem{extradata} Cuts for 7--10\,keV interval of recoil energy
were frozen before data were unmasked.
The decision to report these data was made after the unmasking.
Inclusion of this interval increases the expected background by
$0.1\pm0.02$(stat) events, and lowers the limit on WIMP-nucleon cross
section only for WIMP masses in the 8--11\,\gev\ interval.  No
candidate events were found in the 7--10\,keV interval of recoil energy.
\end{thebibliography}
\end{document}